\documentstyle[prd,aps,preprint,floats]{revtex}
\tightenlines
\input epsf.tex
\def\DESepsf(#1 width #2){\epsfxsize=#2 \epsfbox{#1}}
\def\Journal#1#2#3#4{{#1} {\bf #2}, #3 (#4)}

\def\NPB{{\em Nucl. Phys.} B}
\def\NPA{{\em Nucl. Phys.} A}
\def\PLB{{\em Phys. Lett.}  B}
\def\PRL{\em Phys. Rev. Lett.}
\def\PRD{{\em Phys. Rev.} D}
\def\PRA{{\em Phys. Rev.} A}
\def\ZPC{{\em Z. Phys.} C}

\def\PR{\em Phys. Rep.}

\def\JPG{{\em J. Phys.} G}
\begin{document}

\draft
\preprint{\hbox{CTP-TAMU-19-01}}
\title{SUSY Phases, the Electron Electric Dipole Moment and the Muon Magnetic
Moment} 

\author{R. Arnowitt, B. Dutta and Y. Santoso }

\address{
Center For Theoretical Physics, Department of Physics, \\
Texas A$\&$M University, College Station TX 77843-4242}
\date{June, 2001} 
\maketitle
\begin{abstract}
The electron electric dipole moment ($d_e$) and the muon magnetic 
moment anomaly ($a_{\mu}$) recently observed at BNL are analyzed within the
framework of SUGRA models 
with CP violating phases at the GUT scale. It is seen analytically that 
even if $d_e$ were zero, there can be a large Bino mass phase (ranging from 
0 to $2 \pi$) with a corresponding large $B$ soft breaking mass phase (of size 
$\stackrel{<}{\sim} 0.5$ with sign fixed by the experimental sign of $a_{\mu}$).
The dependence of 
the $B$ phase on the other SUSY parameters, gaugino mass $m_{1/2}$, $\tan
\beta$, $A_0$, is examined.
The lower bound of $a_{\mu}$ determines the upper bound of 
$m_{1/2}$. 
It is shown analytically how the existence of a non-zero Bino phase reduces 
this upper bound (which would correspondingly lower the SUSY mass spectra). 
The experimental upper bound on $d_e$ determines the range of allowed phases, 
and the question of whether the current bound on $d_e$ requires any fine 
tuning is investigated. At the electroweak scale, the phases have to be 
specified to within a few percent. At the GUT scale, however, the $B$ phase 
requires fine tuning below the $1 \%$ level over parts of the parameter space 
for low $m_{1/2}$, and if the current experimental bound on $d_e$ were reduced
by 
only a factor of $3-4$, fine tuning below $1 \%$ would occur at both the 
electroweak and GUT scale over large regions of the parameter space. All 
accelerator constraints ($m_h > 114$ GeV, $b \rightarrow s \gamma$, etc.) and
relic density 
constraints with all stau-neutralino co-annihilation processes are included 
in the analysis.
\end{abstract}

\newpage
\section{Introduction}
The role that CP violation plays in elementary particle physics and how it 
relates to current theory, still remains after many years unclear. In the 
Standard Model (SM), CP violation is accommodated by inserting a single 
phase into the CKM matrix, and there are now under way a number of 
experiments to test the validity of this idea~\cite{CPexp}. In supersymmetry,
the CKM  phase can also exist, but it is possible to have additional phases 
appearing in the soft breaking masses.  From the viewpoint of string 
theory, CP violating phases are a natural occurrence.  Thus in 10 or 11 
dimensional M-theory models with six dimensions compactified on a 
Calabi-Yau manifold, the Kahler potential and Yukawa matrices are 
represented by integrals over the Calabi-Yau space, and since this is a 
complex manifold, it is not surprising that CP violating phases can arise. 
With supersymmetry breaking, phases could arise also in the soft breaking 
masses. However, M-theory is not sufficiently developed to determine the 
details of such effects, and it is useful to have phenomenological 
constraints as a guide to where the fundamental theory may be.

As is well known, the existance of CP violating phases in the SUSY soft 
breaking masses leads to electric dipole moments (EDMs) for the electron 
and neutron, and the smallness of the current experimental bounds on these puts 
severe constraints on the parameter space. One may satisfy these 
constraints by assuming that the phases are non-zero but small (i.e. 
$O(10^{-2})$) or that the SUSY masses are large (i.e. $\stackrel{>}{\sim} O(1\,
{\rm TeV})$). The electron 
EDM (eEDM) arises from two diagrams, one involving the chargino-sneutrino 
intermediate state, and one involving the neutralino- selectron 
intermediate state. More recently it was suggested that a cancellation might 
occur between these two for relatively large phases (i.e.
$O(10^{-1})$)  over 
a reasonably large range of parameters~\cite{cancel}, and there has been a
large number of analyses based on this 
idea~\cite{largephase,aad,199HgEDM,199HgEDM1}. Diagrams similar
to the above 
also enter into the muon $g- 2$, and the recently reported 2.6 $\sigma$ 
deviation from the Standard Model value for that quantity~\cite{BNL} has placed 
significant bounds (at the $95 \%$ C.L.) on the allowed SUGRA parameter
space~\cite{adhs}. It is thus natural to ask whether both constraints can be 
phenomenologically realized, and initial discussions of this have been
made~\cite{g2edm}.

The minimal supersymmetric standard model (MSSM) has over 40 independent 
phases, and so cannot make significant theoretical predictions on this 
question. We use here instead supergravity (SUGRA) GUT models with gravity 
mediated supersymmetry breaking~\cite{sugra} and R-parity
invariance~\footnote{Two
alternate SUGRA models are anomaly mediated~\cite{anomaly} and gauge
mediated~\cite{gauge} 
soft breaking. The former appears to have difficulty in satisfying the 
Brookhaven E821 bounds on $g - 2$ when no SUSY CP phases are
present~\cite{g-2ano}, 
while the latter does not appear to have a satisfactory dark matter 
candidate~\cite{gmnoDM}.}. Such models 
have become quite predictive, in part due to the fact that they 
automatically include the LEP results on grand unification, and because 
radiative breaking of $SU(2) \times U(1)$ implies that the SUSY soft breaking 
masses are of electroweak size. Thus the general size of both $g - 2$ and the 
EDMs (which depend on the size of the SUSY masses) are restricted. In addition,
such 
models have a natural candidate for dark matter, the lightest neutralino 
($\tilde{\chi}_1^0$), and the condition that the relic density of neutralinos be
in accord with the allowed range from astronomical measurements also puts 
important constraints on the parameter space.

Previous analyses of the EDMs within the framework of SUGRA GUT models has 
shown that a new fine tuning problem arises at the GUT scale for $\theta_{B_0}$
(the phase of B, the bilinear soft breaking mass) when $\tan \beta$ 
gets large~\cite{aad}. Further, the discussion of the muon $g - 2$ have shown
that $\tan \beta$ 
is greater than 5 ( and very possibly greater than 10). In carrying out our 
analysis, then, we put a constraint  on the parameter space that 
$\Delta \theta_{B_0} /\theta_{B_0} > 0.01$, where $\Delta \theta_{B_0}$ is the
allowed range that will satisfy the experimental bounds on the EDMs. In order to
carry out a complete analysis, however, it is necessary to include all the 
accelerator constraints (i. e. that $m_h > 114$ GeV (where $h$ is the light 
Higgs boson), the CLEO bound on the $b \rightarrow s  \gamma$ decay, etc.) as
well as 
the full analysis of the relic density including all the stau-neutralino 
co-annihilation channels. (The details of these were discussed
in~\cite{coan,coan1}.)

Both the electron EDM and the muon $g - 2$ can be calculated from two types 
of diagrams, one with the intermediate chargino and sneutrinos states, and one 
with the intermediate neutralino and slepton states. If one assumes universal 
soft breaking in the first two generations, then the eEDM and the SUSY deviation
of the muon magnetic moment from its Standard Model value, $a_{\mu}^{\rm SUSY}$
($a_{\mu} =(g_{\mu} - 2)/2$), 
can both be calculated from a single amplitude for these diagrams, 
the former being related to the imaginary part, and the latter to the real 
part. In the following we will use the symbol $a_{\mu}$ to mean the deviation of
the muon $(g_{\mu} - 2)/2$ from its Standard Model value.

As is well known, for large $\tan \beta$, $a_{\mu}$ is dominated by the
chargino 
diagram~\cite{b14,b14a} (the neutralino diagrams being generally quite small), while 
the strong experimental constraints on the eEDM require a near cancellation 
between the neutralino and chargino diagrams~\cite{cancel}. In order to see how
this 
can come about when CP violating phases are present, we calculate in the 
Appendix the leading large $\tan\beta$ terms. We discuss analytically in Sec. 2 
how both experimental constraints can naturally be satisfied at the 
electroweak scale. In Sec. 3 we examine in detail numerically the combined
experimental 
constraints of the eEDM and the Brookhaven E821 $a_{\mu}$ experiment on  the
SUSY parameter 
space for a variant of the mSUGRA model where the magnitudes of the soft
breaking masses are universal at $M_G$ but the phases are arbitrary.  
The experimental lower bound on $a_{\mu}$ puts an upper bound on 
the gaugino mass $m_{1/2}$, and this upper bound is generally reduced when CP 
violating phases are present. The Higgs mass lower bound and the $b  \rightarrow
s \gamma$ constraint generally put a lower bound on $m_{1/2}$. Imposing the
fine tuning constraint at $M_{\rm GUT}$ generally increases that bound when CP
violating phases are present and also limits the range of the CP violating
phases. Thus the two experiments interact to further restrict the SUSY 
parameter space when 
CP violating phases are large. In Sec. 4 we give some concluding remarks.

\section{Muon Magnetic Moment and Electron EDM}

We consider here supergravity models which are a generalization of mSUGRA 
to include the possibility of phases in the soft breaking masses at the GUT 
scale $M_G \cong 2 \times 10^{16}$ GeV. The magnitude of the soft breaking
masses are 
still assumed to be universal, but phases are not necessarily universal. 
Thus the SUSY parameters at $M_G$ are $m_0$ (the scalar soft breaking mass), 
$m_{1/2i} = |m_{1/2}| \exp(i \phi_i)$ i = 1,2,3 (the gaugino masses for the
$U(1)$, $SU(2)$, $SU(3)$ groups), $A_0 = |A_0| \exp(i \alpha_0)$ (cubic soft
breaking mass), 
$B_0 = |B_0| \exp(i \theta_{B_0})$ (quadratic soft breaking mass), and $\mu_0 = 
|\mu_0| \exp(i \theta_{\mu})$ (the Higgs mixing parameter in the
superpotential). 
The model therefore depends on five magnitudes and six phases. However, one 
can always set one of the gaugino phases to zero, and we chose here $\phi_2 = 
0$.

The renormalization group equations (RGEs) allow one to evaluate the 
parameters at the electroweak scale. To one loop order, the $\phi_i$ and 
$\theta_{\mu}$ do not run. Further, radiative breaking of $SU(2) \times U(1)$ at
the 
electroweak scale allows one to eliminate $\theta_{\mu}$ in terms of $\theta_B$,
the $B$ phase at the  
electroweak scale, and determine $|\mu|$ and $|B|$ in terms of
the 
other parameters and $\tan \beta = |\langle H_2 \rangle / \langle H_1 \rangle
|$.
Thus with a convenient 
choice of Higgs VEV phases, the minimization of the Higgs potential
yields~\cite{b16}:
\begin{eqnarray}
\theta_{\mu} =  - \theta_B  +  f(-\theta_B+\alpha_q, -\theta_B+\alpha_l) \\
|\mu|^2 = \frac{m_1^2-\tan^2 \beta \; m_2^2}{\tan^2 \beta -1} ; \;\;\;\;\;   
|B| = \frac{1}{2} \sin 2\beta \; \frac{m_3^2}{|\mu|} 
\end{eqnarray}
where $\alpha_q$ and $\alpha_l$ are the quark and lepton phase of $A_q$ and
$A_l$ at the electroweak scale, $m_i^2 = m_{H_i}^2  + \Sigma_i$ ($i=1,2$), 
$m_3^2 = 2|\mu|^2 + m_1^2 + m_2^2$,
and $m_{H_i}$ are the Higgs running masses at the electroweak scale. $\Sigma_i =
dV_1/dv_i^2$  where $V_1$ is the one loop contribution to the Higgs potential, 
and $v_i = |\langle H_i \rangle |$. There remain therefore four real parameters
which we 
take to be $m_0$, $|A_0|$, $|m_{1/2}|$ and $\tan \beta$, and four phases
$\theta_{B_0}$, $\alpha_0$, 
$\phi_1$ and $\phi_3$. In this paper we examine only the electron electric
dipole moment\footnote{While there has been considerable progress in calculating
the neutron 
EDM~\cite{nEDM} and $^{199}$Hg EDM~\cite{199HgEDM}, there remain still
significant hadronic 
uncertainties, in contrast to the clear calculation of the eEDM. Further, 
the $\phi_3$ phase can always be adjusted to satisfy the  neutron EDM, and the 
effect of the $^{199}$Hg EDM would then only result in reducing the remaining 
parameter space that we find from the eEDM and the muon $g - 2$ given here.}
(eEDM), and so our results are only very weakly dependent on $\phi_3$. 
Also, since we are dealing only with first and second generation sleptons, 
there is only a weak dependence on $\alpha_0$. Thus the two important phases 
are $\theta_{B_0}$ and $\phi_1$. The parameter range that we examine is:
\begin{eqnarray}
|m_0| < 1 \;{\rm TeV}; \;\;\;\;  |m_{1/2}| < 1 \; {\rm TeV}; \;\;\;\;  
|A_0 / m_{1/2}| < 4      
\end{eqnarray}
\begin{eqnarray}
\tan \beta  \leq 40 
\end{eqnarray}

As discussed above, both $a_{\mu}$ and $d_e$ can be obtained from the same
complex 
amplitude  $A$ (assuming universal soft breaking in the first two 
generations). One has then
\begin{eqnarray}
a_{\mu} = - \frac{\alpha}{4 \pi \sin^2 \theta_W} \, m_{\mu}^2 \, {\rm Re}[A] \\ 
d_e/e = - \frac{\alpha}{8 \pi \sin^2 \theta_W} \, m_e \, {\rm Im}[A] 
\end{eqnarray}

The amplitude $A$ is defined in the Appendix. For the case where there are no 
CP violating phases, the experimental $a_{\mu}$ data~\cite{BNL} favors large
$\tan \beta$~\cite{adhs}. In order to see semi-quantitatively the nature of the
cancellations 
needed to make $d_e$ small, the leading terms for large $\tan\beta$ were 
obtained in the 
Appendix when $\mu^2 \gg M_W^2$ (which is almost always the case for the 
mSUGRA  parameter space). From Eqs.~(A18, A25) we find for the neutralino and 
chargino diagrams
\begin{equation}
a_{\mu} = a(\tilde{\chi}^{\pm}) + a(\tilde{\chi}^0) 
\end{equation}
where
\begin{eqnarray}
a(\tilde{\chi}^{\pm}) &=& C_{\mu}(\tilde{\chi}^{\pm}) \left[
\frac{|\mu|^2}{|\mu|^2-\tilde{m}_2^2} F_1 -
\frac{\tilde{m}_2^2}{|\mu|^2-\tilde{m}_2^2} F_2 \right] \cos \theta_{\mu} \\
a(\tilde{\chi}^0) &=& C_{\mu}(\tilde{\chi}^0) \left[ \left\{ \left(
\frac{|\mu|^2}{m_{\tilde{u}_L}^2-m_{\tilde{u}_R^2}}-
\frac{|\mu|^2}{|\mu|^2-\tilde{m}_1^2} \right) G_{11} \right. \right. \nonumber
\\  
&& \left. - \left( \frac{|\mu|^2}{m_{\tilde{u}_L}^2-m_{\tilde{u}_R^2}}-
\frac{1}{2} \frac{|\mu|^2}{|\mu|^2-\tilde{m}_1^2} \right) G_{21} \right\}
\cos(\theta_{\mu} + \phi_1) \nonumber \\
&& \left. - \frac{1}{2 \tan^2 \theta_W}
\frac{|\tilde{m}_1|}{\tilde{m}_2} \frac{|\mu|^2}{|\mu|^2-\tilde{m}_2^2} \left\{
\left( G_{22} - \frac{1}{2} \left( \frac{\tilde{m}_2}{|\mu|} \right)^2 G_{23}
\right) \cos \theta_{\mu} - \frac{1}{2} \frac{\tilde{m}_2}{|\mu|} G_{23} 
\right\} \right] 
\end{eqnarray}
where
\begin{equation}
C_{\mu}(\tilde{\chi}^{\pm}) = \frac{\alpha \, m_{\mu}^2 \tan \beta}{4 \pi 
\tilde{m}_2 |\mu| \sin^2 \theta_W} 
\end{equation}
and $C_{\mu}(\tilde{\chi}^0) = \tan^2 \theta_W (\tilde{m}_2/|\tilde{m}_1|)
C_{\mu}(\tilde{\chi}^{\pm})$. The form factors $F_i = 
F(m_{\tilde{\nu}}^2/m_{\tilde{\chi}_i^{\pm}}^2)$ and $G_{ki} =
G(m_{\tilde{\mu}_k}^2/m_{\tilde{\chi}_i^{0}}^2)$ are defined in Eqs.~(A8, A12),
and $|\tilde{m}_1| \cong 0.4 m_{1/2}$, $\tilde{m}_2 \cong 0.8 m_{1/2}$. (Our
states are labeled such that e.g. $m_{\tilde{\chi}_i^0} < m_{\tilde{\chi}_j^0}$
for $i < j$.)

A similar decomposition of the electron electric dipole moment, 
$d_e/e = d(\tilde{\chi}^{\pm}) +  d(\tilde{\chi}^0)$,  gives:
\begin{eqnarray}
d(\tilde{\chi}^{\pm}) &=& -D_e(\tilde{\chi}^{\pm}) \left[
\frac{|\mu|^2}{|\mu|^2-\tilde{m}_2^2} F_1 -
\frac{\tilde{m}_2^2}{|\mu|^2-\tilde{m}_2^2} F_2 \right] \sin \theta_{\mu} \\
d(\tilde{\chi}^0) &=& -D_e(\tilde{\chi}^0) \left[ \left\{ \left(
\frac{|\mu|^2}{m_{\tilde{u}_L}^2-m_{\tilde{u}_R^2}}-
\frac{|\mu|^2}{|\mu|^2-\tilde{m}_1^2} \right) G_{11} \right. \right. \nonumber
\\  
&& \left. - \left( \frac{|\mu|^2}{m_{\tilde{u}_L}^2-m_{\tilde{u}_R^2}}-
\frac{1}{2} \frac{|\mu|^2}{|\mu|^2-\tilde{m}_1^2} \right) G_{21} \right\}
\sin(\theta_{\mu} + \phi_1) \nonumber \\
&& \left. - \frac{1}{2 \tan^2 \theta_W}
\frac{|\tilde{m}_1|}{\tilde{m}_2} \frac{|\mu|^2}{|\mu|^2-\tilde{m}_2^2} 
\left( G_{22} - \frac{1}{2} \left( \frac{\tilde{m}_2}{|\mu|} \right)^2 G_{23}
\right) \sin \theta_{\mu} 
\right]        
\end{eqnarray}
where $D_e = (m_e/2 m_{\mu}^2) C_{\mu}$.

In the case where all phases are zero, it is well known that the chargino 
contribution to $a_{\mu}$ dominates over the neutralino  diagram even though
the 
front factors $C_{\mu}(\tilde{\chi}^{\pm})$ and $C_{\mu}(\chi^0)$ 
are of comparable size. This can be 
understood from Eqs.~(8) and (9) in the following way. The second term in 
Eq.~(8), coming from the heavy chargino, $\tilde{\chi}_2^{\pm}$, cancels about
$30 \%$ of
the 
light chargino contribution.  In a similar fashion, the second term in Eq.~(10)
coming from the heavy smuon , $\tilde{\mu}_2$, contribution cancels part of
the leading term. However, due to the slowly varying nature of the form 
factors $G_{11}$ and $G_{21}$, about $75 \%$ of the leading $G_{11}$ term
is canceled, reducing $a(\tilde{\chi}^0)$ significantly. 
The remaining two terms arising from the heavy neutralinos,
$\tilde{\chi}^0_{2,3,4}$ 
are generally small (and there can be cancellations also between these 
terms). In contrast, when the CP violating phases are not zero and an 
electric dipole moment exits, the neutralino and chargino contributions 
must be of nearly equal size and opposite sign so that $d_e$ be greatly 
suppressed to be in accord with experiment. This change in the relative 
sizes of the two terms must be brought about by the  sine factors in 
Eqs.~(11, 12). 
Simultaneously, the corresponding cosine factors in Eqs.~(8, 9) 
will modify the predictions of $a_{\mu}$. This then leads to two questions:

(1) Can the experimental constraints on $d_e$ and $a_{\mu}$ (and of course all
other 
experimental constraints) be simultaneously satisfied with ``large'' phases, 
i. e.  of O($10^{-1}$)?

(2) If large phases are possible (and $d_e$ is very small) can the 
experimental constraints be satisfied without undue fine tuning of the phases?

Question (2) divides into two parts: (2a) Is there undue fine tuning of 
phases needed to satisfy the experimental constraints at the electroweak 
scale, and (2b) Is there undue fine tuning of the phases needed at the GUT
scale? 
In the following, we define the fine tuning parameter $R(\phi)$ for any phase 
$\phi$ by
\begin{equation}
R(\phi) \equiv \frac{\phi_2 - 
\phi_1}{\frac{1}{2}(\phi_2 + 
\phi_1)} \equiv \frac{\Delta \phi}{\phi_{\rm av}}
\end{equation}
where $\phi_2$ and $\phi_1$ are the upper and lower value of $\phi$ when the 
experimental constraints on $d_e$ and $a_{\mu}$ are both satisfied. We use here
the current bound on $d_e$~\cite{de}:
\begin{equation}
|d_e| < 4.3 \times 10^{-27} \; e \; {\rm cm} 
\end{equation}
and the 2 std range for $a_{\mu}$~\cite{BNL}:
\begin{equation}
11 \times 10^{-10} < a_{\mu} < 75 \times 10^{-10}
\end{equation} 
In the following we will assume that no fine tuning less than $1 \%$ be 
allowed, i. e.
\begin{equation}
R(\phi) > 0.01 
\end{equation}

Within this framework, we find that Question (1) can be answered 
affirmatively, i. e. cancellations between the neutralino and chargino 
diagrams in $d_e$ can indeed be satisfied with large phases. Further, the 
answer to Question 2(a) is that no undue fine tuning of the electroweak 
parameters is necessary to achieve the suppression of $d_e$. However, we will 
see that this is not the case at the GUT scale, where the simultaneous 
requirements of electroweak radiative breaking and the experimental $d_e$ 
constraint leads to significant fine tuning, and if Eq.~(16) were imposed, 
a large amount of the parameter space is eliminated.

In a GUT theory, the fundamental parameters are specified at $M_{\rm GUT}$, and
the 
consequences at the electroweak scale are obtained from the renormalization 
group equations (RGEs). These GUT parameters are presumably to be 
determined at some future time by a more fundamental theory (e. g. string 
theory), and so fine tuning at the GUT scale may imply a significant 
theoretical problem (while fine tuning at the electroweak scale would be an 
acceptable theoretical consequence of the RGEs). Of course, what level of 
fine tuning is acceptable is somewhat a matter of choice, and we view
Eq.~(16) only as a reasonable benchmark to consider.

In Sec. 3 below, we will consider these results in detail quantitatively. 
We give here an analytic discussion of how they arise. To show that large 
phases are easily achievable, we write $d_e$ in the form
\begin{equation}
d_e/e = - D_e(\tilde{\chi}^{\pm}) A [\sin(\theta_{\mu}) + a 
\sin(\theta_{\mu}+ 
\phi_1)]
\end{equation}
where the coefficients $A$ and $a$ can be read off from Eqs.~(11, 12). In the 
extreme case where $d_e = 0$, Eq.~(17) and Eq.~(1) imply
\begin{equation}
\tan \theta_B = \frac{a \sin \phi_1}{1 + a \cos \phi_1} 
\end{equation}
where we have neglected the small 1-loop correction in Eq.~(1). The fact 
that the chargino diagram dominates over the neutralino diagram implies 
that $a < 1$, and detailed numerical calculations show that $a \sim 0.2 - 0.4$
for 
much of the SUSY parameter space. Thus as $\phi_1$ varies from 0 to $2 \pi$,
over 
most of the parameter space $|\theta_B|$ will be large (rising to a maximum of 
about 0.5) even though $d_e$ has been set to zero!

We next consider the effects of the CP violating phases on $a_{\mu}$. From 
Eqs.~(8, 9), we have
\begin{equation}
a_{\mu} = C_{\mu}(\tilde{\chi}^{\pm}) A [\cos(\theta_{\mu}) + 
a \cos( \theta_{\mu} + \phi_1) + b]
\end{equation}
where $b$ is the term in Eq.~(9) independent of the phases. In general, $b$ is 
quite small, and  we will neglect it in the following. (Of course, in the 
numerical calculations in Sec. 3 all such effects as well as the loop 
corrections are considered.) Using Eq.~(18) to eliminate $\theta_B$, we obtain
\begin{equation}
a_{\mu} = \pm \; a_{\mu}(0) \; Q(a,\phi_1) 
\end{equation}
where $a_{\mu} (0)$ is the value of $a_{\mu}$ with zero phases,
\begin{equation}
Q = \frac{[1 + 2 a \cos \phi_1 + a^2]^{1/2}}{(1 + a)} \leq 1
\end{equation}
The $\pm$ factor in Eq.~(20) is the sign of $\cos \theta_B$. Since
experimentally, 
$a_{\mu} > 0$, this implies
\begin{equation}
\theta_B > 0 \;\;\;\; {\rm for} \;\;\; 0 < \phi_1 < \pi; \;\;\;\;
\theta_B < 0 \;\;\;\; {\rm for} \;\;\; \pi < \phi_1 < 2 \pi 
\end{equation}

The two branches of Eq.~(22) are symmetric, and in the following we consider
the 
$\theta_B > 0$ branch. The factor $Q$ in Eq.~(21) reduces the theoretically 
expected size for $a_{\mu}$. Since the experimental lower bound on $a_{\mu}$
implies an 
upper bound on $m_{1/2}$~\cite{adhs}, the $Q$ factor due to the phases will reduce this 
upper bound, further restricting the allowed SUSY parameter space. However, 
since in general $Q \stackrel{>}{\sim} 0.5$, this reduction will still be
consistent with all 
experimental data, and the effect of the SUSY CP violating phases does not 
qualitatively change the fit to the data for $a_{\mu}$ that was obtained
assuming no CP violating phases.

We can also estimate how much fine tuning is needed in the phases to 
satisfy the experimental bound of Eq.~(14). Thus let $\Delta \theta_B$ be the 
change in $\theta_B$ for a fixed value of $\phi_1$ as $d_e/e$ varies from $-4.3
\times 10^{-27}$ cm to $+4.3 \times 10^{-27}$ cm.  Characteristically, the
factor $D(\tilde{\chi}^{\pm}) A$ in Eq.~(17) is numerically about 100 times 
the current upper bound on $d_e$ (e.g. for 
$m_{1/2} = 480$ GeV ($m_0 = 118$ GeV), $|\mu| = 690$ GeV and $\tan \beta = 15$,
this factor is 
$4.1 \times 10^{-25}$ cm.)  Considering then the variation of $\theta_B$ in 
Eq.~(17) 
one finds
\begin{equation}
2 \times 10^{-2} \cong \Delta \theta_B \cos(\theta_B) [ 1 + a \cos(\phi_1) + a 
\sin(\phi_1) \tan(\theta_B)]
\end{equation}
Hence using Eq.~(18),
\begin{equation}
 R(\theta_B) \cong \Delta \theta_B/\sin(\theta_B) \cong 2 \times 10^{-2}/(a 
\sin(\phi_1))  
\end{equation}
Thus the allowed range of $\theta_B$ is indeed small, though the condition of 
Eq.~(16) is generally satisfied. A reduction of $d_e$ by a factor of 10 would 
be enough to produce a serious fine tuning problem at the electroweak 
scale. In contrast, we will see below that there is already a significant fine 
tuning problem at the GUT scale even with the current bound on $d_e$.

\section{Detailed Calculations}

In this section, we consider detailed calculations of the effects of the 
experimental constraints involving the eEDM and $a_{\mu}$ which were
analytically 
estimated in Sec.~2. The analysis is done within the framework of the
generalized mSUGRA described in Sec.~2
(though extension to non-universal soft breaking models can easily be 
done). In order to get a clear picture of what constraints on the SUSY 
parameter space arise, it is necessary to simultaneously impose all the 
experimental constraints. Aside from the above, these include the 
following: (1) The LEP Higgs  mass lower bound $m_h > 114$ GeV~\cite{mhexp}.
Since the 
theoretical analysis of $m_h$ still has about a 3 GeV uncertainty, (which 
(conservatively) may be an overestimate) we interpret this in the 
theoretical calculation~\cite{coan} to mean $m_h > 111$ GeV. (2) The $b 
\rightarrow s \gamma$ 
branching ratio constraint. We  take here a 2 std range of the experimental CLEO
data~\cite{bsgexp}
\begin{equation}
1.8 \times 10^{-4} \leq BR(B \rightarrow X_s \gamma) \leq 4.5 \times 10^{-4}
\end{equation}
In the theoretical analysis we 
include the NLO SUSY contribution for 
large $\tan \beta$~\cite{bsgNLO}, as these produce significant effects
(particularly 
since the $a_{\mu}$ lower bound favors larger values of $\tan \beta$). (4) We
include the 
1-loop corrections to the $b$ and $\tau$ masses~\cite{aads}, which are
significant also 
for large $\tan \beta$. (5) All co-annihilation effects are included in the 
relic density calculations. We assume here the range for the neutralino 
cold dark matter to be
\begin{equation}
0.025 < \Omega_{\tilde{\chi}_1^0} h^2 < 0.25 
\end{equation}
The upper bound is consistent with recent Boomerang and Maxima
measurements~\cite{boomerang,maxima}, while the lower bound allows for the
possibility that there may be 
more than one type of dark matter. However, our calculations here are 
insensitive to the precise value of the lower bound, and one would get very 
similar results if the lower bound were raised to 0.05 or 0.1.

The effects of the Higgs mass and $b \rightarrow s \gamma$ bounds is to push the
allowed 
parameter space mostly into the co-annihilation domain of the relic density 
calculation. (Thus only a small part of the allowed parameter space occurs 
at small enough $m_{1/2}$ to lie below the region of co-annihilation, which 
begins at $m_{1/2} \cong 350-400$ GeV.) In SUGRA models, stau - neutralino 
co-annihilation can occur quite naturally. Thus for $m_0 = 0$, the charged 
sleptons lie below the lightest neutralino, $\tilde{\chi}_1^0$, and one must
increase $m_0$ to 
raise their masses so that the $\tilde{\chi}_1^0$ is the dark matter particle. 
Thus as $m_{1/2}$  increases, $m_0$ must be increased in
lock 
step so that the neutralino remains the LSP, and one finds a relatively 
narrow corridor in the $m_0-m_{1/2}$ plane consistent with the relic density 
bound of Eq.~(26) and with the lightest stau  lying above the neutralino. 
The dependence of these corridors on $\tan \beta$ and $A_0$ are shown in Figs.~1
and 2 for the case where there are no SUSY CP violating phases. We see that 
the corridors (which are characteristically 20-30 GeV wide) lie higher for 
larger $\tan \beta$ and larger $|A_0|$. This is because the stau mass decreases 
when these parameters are increased, and so one must raise $m_0$ to keep the 
stau mass greater than the neutralino\footnote{Increasing $\tan \beta$ or making
$A_0$ negative increases the magnitude of the 
L-R term in the stau mass matrix of Eq.~(A3) and hence decreases the 
light stau mass. For $A_0 > 0$, the opposite effect occurs, but also the 
diagonal matrix elements of the mass matrix are reduced, and again the stau 
mass decreases.}. In general, the effect of 
co-annihilation is to determine $m_0$ approximately in terms of $m_{1/2}$ for a 
given $\tan \beta$ and $A_0$. This greatly sharpens the predictions of the
theory.

\begin{figure}[htb]
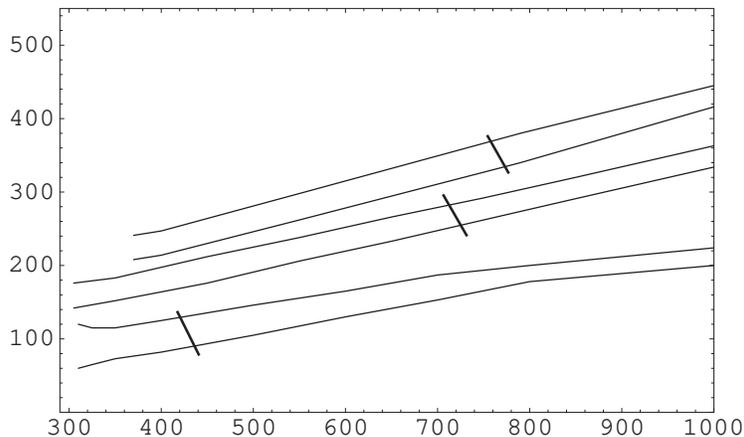

 \centerline{ \DESepsf(edmt16.epsf width 10 cm) }
\caption {\label{fig1}Corridors in the $m_0 - m_{1/2}$ plane allowed by the relic density 
constraint for (bottom to top) $\tan \beta = 10$, 30, 40 for $m_h > 114$ GeV,
$A_0 = 
0$, $\mu > 0$, all phases set to zero. The $\tan \beta = 30$ and 40 corridors
all lie 
in the co-annihilation region, while only the beginning of the $\tan \beta = 10$
corridor is in the non co-annihilation region. The Higgs mass bound 
determines the lower $m_{1/2}$ bound for $\tan \beta = 10$, while both
the 
Higgs mass and the $b \rightarrow s \gamma$ bounds equally produce the lower
bound for 
$\tan \beta = 30$ and
 $b \rightarrow s \gamma$ determines the lower bound for $\tan\beta=$40. The short slanted lines represent the  upper bound on 
$m_{1/2}$ due to the $a_{\mu}$ lower bound, of Eq.~(15).
}
\end{figure}

\begin{figure}[htb]
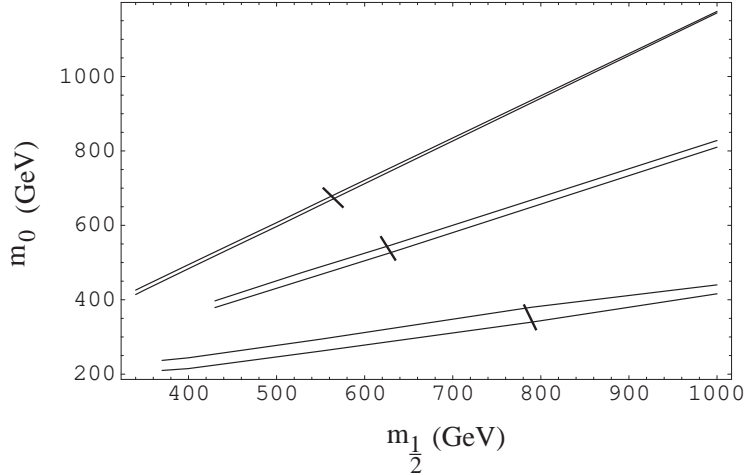

\centerline{ \DESepsf(adhs3.epsf  width 10 cm) }
\caption {\label{fig2} Corridors in the $m_0-m_{1/2}$ plane allowed by the relic density 
constraint for (from bottom to top) $A_0 = 0,\; -2 m_{1/2},\; 4 m_{1/2}$, for
$m_h > 114$ 
GeV, $\mu >0$, all phases set to zero. The lower $m_{1/2}$ bounds for $A_0 = 0,
\;
-2 m_{1/2}$ are due to the $b \rightarrow s \gamma$ constraint, and for $A_0= 4
m_{1/2}$ from the Higgs mass bound~[15].}
\end{figure}

\subsection{Allowed Regions at the Electroweak
Scale for $\theta_B$ and $\phi_1$}

We now examine the effects of having non-zero CP violating phases present, 
and discuss the dependence of the allowed phases on the SUSY parameters. To 
illustrate the phenomena, we consider one low $\tan \beta$ and one high $\tan
\beta$. 
Since for no CP violating phases, one had (at $90 \%$ C.L.) $\tan \beta > 10$,
we 
examine the cases of $\tan \beta = 15$ and $\tan \beta = 40$. Fig.~3 shows the 
corridors allowed for $\theta_B$  (the $B$ phase at the electroweak scale) by
the 
current experimental bounds on $d_e$ for $\tan \beta = 40$, $A_0 =0$ for two
choices 
of the gaugino phase: $\phi_1 = 0.9$ (lower curves) and $\phi_1 = 1.2$ (upper
curves). 
One sees that one gets a significant phase $\theta_B$  of the size expected by 
Eq.~(18), the larger $\phi_1$ allowing a larger $\theta_B$, also in accord with
Eq.~(18). 
Note also that the allowed corridors for $\theta_B$ widens as $m_{1/2}$ 
increases as expected, since the experimental $d_e$ constraint is less severe 
for a heavier mass spectrum.  A similar graph is shown in Fig.~4 for $\phi_1 = 
0.9$ (upper curves), $\phi_1 = 3.4$ (lower curves). 
$\theta_B$ turns negative for $\phi_1$ 
in the third quadrant, again as expected from Eq.~(22).

We note in all these curves, the upper bound on $m_{1/2}$ (due to the lower 
bound on $a_{\mu}$) is reduced compared to the case when the CP violating
phases 
are zero. (Then the upper bound is $m_{1/2} = 790$ GeV for $\tan \beta = 40$,
$A_0 = 0$~\cite{adhs}.) This is due to the 
phase $\phi_1$ in the $Q$ factor in Eq.~(20). $Q$ is smallest when $\phi_1$ is
near $\pi$, as 
can be seen in explicitly in Fig.~4 for $\phi_1 = 3.4$

\begin{figure}[htb]
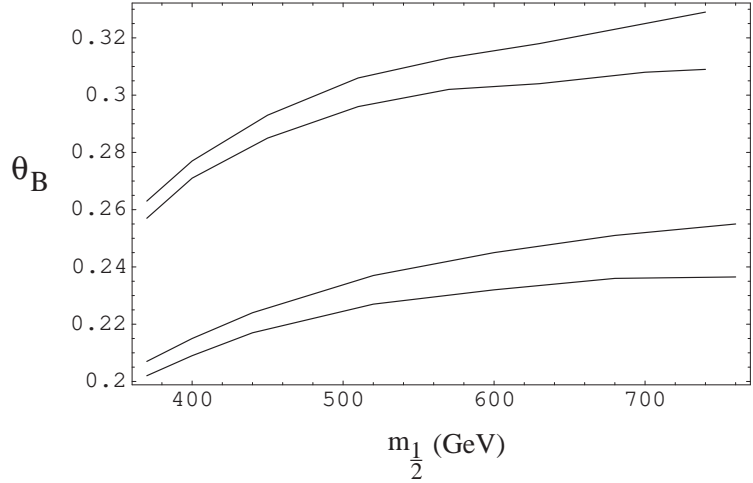

\centerline{ \DESepsf(edmt3.epsf width 10 cm) }
\caption {\label{fig3} 
Regions allowed for $\theta_B$ by the experimental constraint on $d_e$
as a 
function of $m_{1/2}$ for $\tan \beta = 40$, $A_0 = 0$, for $\phi_1 = 0.9$
(lower curves) and $\phi_1 = 1.2$ (upper curves).
}
\end{figure} 

\begin{figure}[htb]
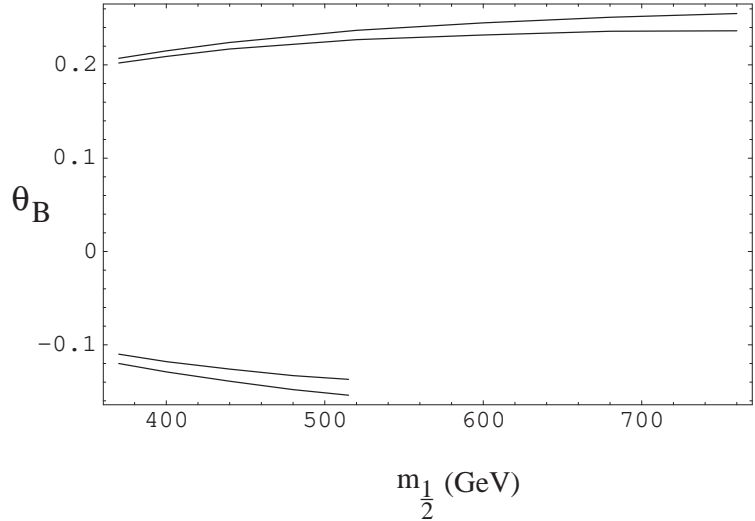

\centerline{ \DESepsf(edmt12.epsf width 10 cm) }
\caption {\label{fig4} 
Same as Fig.~3 for $\phi_1 = 0.9$ (upper curves) and $\phi_1 = 3.4$
(lower curves).
}
\end{figure}

We consider next the $\tan \beta$ dependence of the allowed region for
$\theta_B$. 
This arises due to an indirect $\tan \beta$ dependence in the parameter $a$ of 
Eqs.~(17,18). As seen from Fig.~2, coannihilation determines 
$m_0$ in terms of $m_{1/2}$, and this $m_0$ increases with $\tan\beta$, 
changing the ratio of neutralino to chargino diagram differentially. 
Fig.~5 shows the allowed region for $\tan \beta = 15$,
and 
$A_0 = 0$, $\phi_1 = 1.2$ (upper curves) and $\phi_1 = 0.9$ (lower curves). We
see that 
one can get considerably larger values of $\theta_B$ at lower $\tan \beta$,
though 
the upper bound on $m_{1/2}$ due to the 
lower bound on $a_{\mu}$ is considerably reduced at low $\tan \beta$. 
The $A_0$ dependence is exhibited in Fig.~6, where
the 
allowed corridor for $\theta_B$ is plotted for $\tan \beta = 40$, $\phi_1 = 0.9$
and $A_0 
= 0$ (upper curves), $|A_0| = 2 m_{1/2}$, $\alpha_0 = 0.5$ (lower curves).
Increasing the magnitude of 
$A_0$ increases the value of $m_0$ (by Fig.~2) and reduces the size of
$\theta_B$.

\begin{figure}[htb]
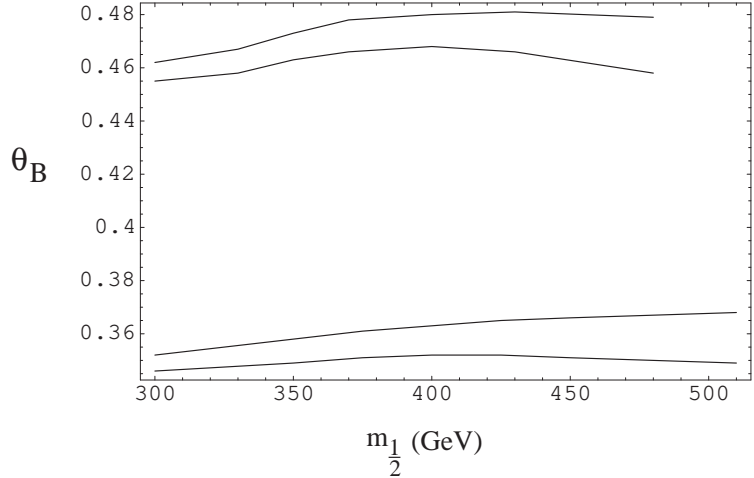

\centerline{ \DESepsf(edmt11.epsf width 10 cm) }
\caption {\label{fig5} 
Same as Fig.~3 for $\tan \beta = 15$. The allowed regions terminate at 
low $m_{1/2}$ due to the $m_h$ constraint.}
\end{figure}

\begin{figure}[htb]
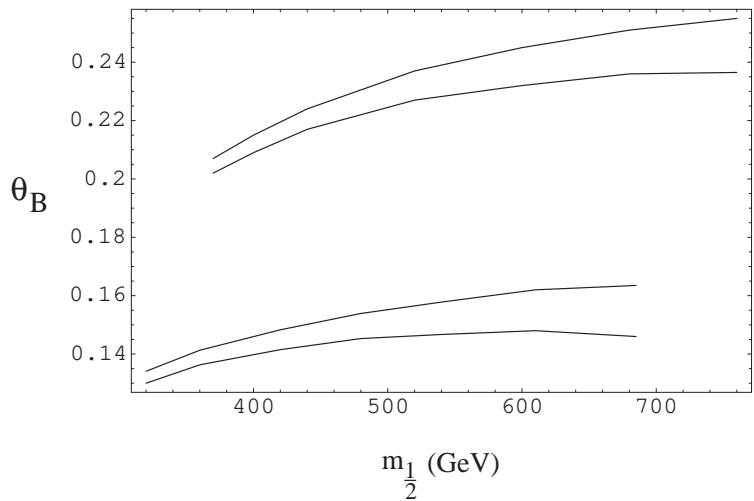

\centerline{ \DESepsf(edmt5.epsf width 10 cm) }
\caption {\label{fig6} 
Allowed region for $\theta_B$  for $\tan \beta = 40$ for $A_0 = 0$ (upper
curves) 
and $|A_0| = 2 m_{1/2}$, $\alpha_0 = 0.5$ (lower curves).}
\end{figure}

\subsection{Fine Tuning at the Electroweak Scale}

The above analysis shows that the $d_e$ experimental constraint allows
$\theta_B$ 
to be $O(10^{-1})$ for a wide region of parameter space, and $\phi_1$ can range 
widely, i.e. from 0 to $2\pi$. We next investigate whether the smallness of 
the upper bound on $d_e$ requires any fine tuning to maintain this constraint. 
In Fig.~7 we plot the fine tuning parameter $R$ of Eq.~(13) for $\theta_B$ for 
$\tan \beta = 40$, $A_0 = 0$ for $\phi_1 = 0.9$ (upper curve) and $\phi_1 = 1.2$
(lower 
curve). One sees that $R(\theta_B)$ is small, i.e. a few percent, but satisfies 
the criteria $R > 0.01$ for the entire range. The size of  $R(\theta_B)$ is 
consistent with what was expected from the analytic analysis of Eq.~(24), 
and increases as $\phi_1$ decreases also as expected. Fig.~8 shows a similar 
plot for $\tan \beta = 15$. Since $\theta_B$ is larger here (see Fig.~5),
$R(\theta_B)$ 
is smaller, but still within the acceptable range. Fig.~9 shows $R(\phi_1)$ for
$\tan \beta = 40$, $A_0 = 0$ for $\theta_B =0.2$
(upper 
curve), and $\theta_B = 0.3$ (lower curve). Again $R(\phi_1) > 0.01$ for the
entire 
range of $m_{1/2}$, and is generally larger than $R(\theta_B)$ (by about a
factor of 
$\tan \phi_1/\phi_1$).

\begin{figure}[htb]
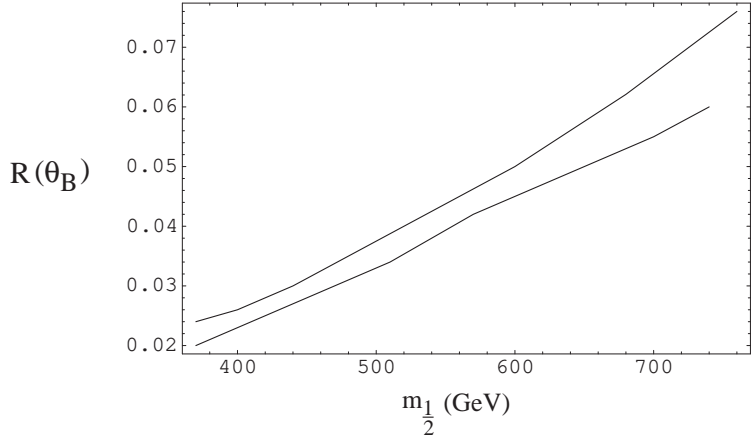

\centerline{ \DESepsf(edmt7.epsf width 10 cm) }
\caption {\label{fig7} 
$R(\theta_B)$ as a function of $m_{1/2}$ for $\tan \beta = 40$, $A_0 =
0$, $\phi_1 = 
0.9$ (upper curve), $\phi_1 = 1.2$ (lower curve).}
\end{figure}
                                                                        
\begin{figure}[htb]
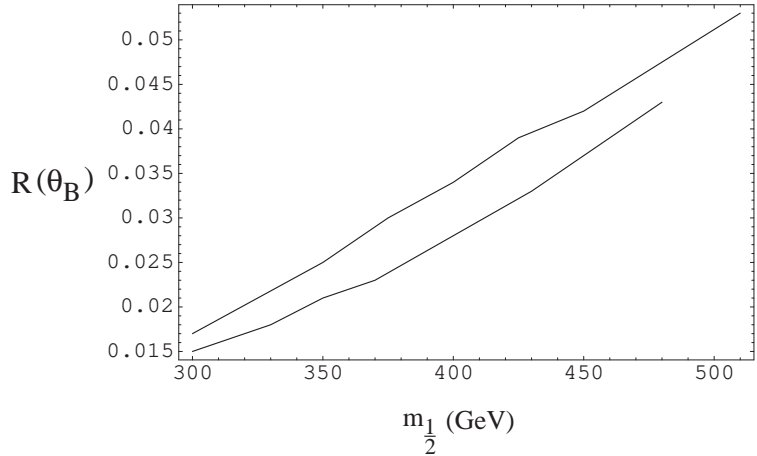

 \centerline{ \DESepsf(edmt8.epsf width 10 cm) }
\caption {\label{fig8}
Same as Fig.~7 for $\tan \beta = 15$.}
\end{figure}

\begin{figure}[htb]
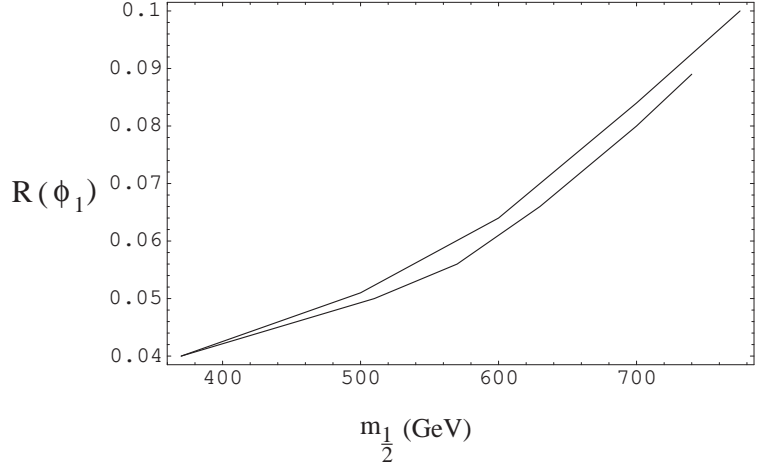

\centerline{ \DESepsf(edmt13.epsf width 10 cm) }
\caption {\label{fig9} 
$R(\phi_1)$ as a function of $m_{1/2}$ for $\tan \beta = 40$, $A_0 = 0$
and $\theta_B = 
0.2$ (upper curve), $\theta_B = 0.3$ (lower curve).}
\end{figure}

We see from the above results that the current experimental bound on $d_e$ can 
be accommodated with large phases. Further, for most of the parameter 
space, while the range of the allowed phases at the electroweak scale 
required to accommodate the bound on $d_e$ is small, no fine tuning below $1 \%$
is 
needed. We will see, however, a more serious fine tuning problem develops 
at the GUT scale.

\subsection{Fine Tuning at the GUT Scale}

We now examine what parameters are fine tuned at the GUT scale to achieve 
the experimental EDM bounds. From Eq.~(1), we see since the loop correction 
is small, that $\theta_{\mu} \cong -\theta_B$, and since we have seen that
$\theta_B$ does 
not need excessive fine tuning, the same can be said for $\theta_{\mu}$.
Further, 
since $\theta_{\mu}$ does not run with the RGE, its value at the GUT scale  
is the same as at the electroweak scale and hence no fine tuning is needed at
$M_G$. A
similar result holds for $\phi_1$, which also does not 
run with the RGE. However, as has been previously pointed out~\cite{aad},
matters 
are different for the $B$ phase at $M_G$, $\theta_{B_0}$, and we review 
briefly the 
discussion given there\footnote{This analysis differs from that given
in~\cite{b30}
which does not take into 
account the possibility of cancellations in $d_e$ between the neutralino and 
chargino diagrams. In fact the discussion in~\cite{b30} sets the phase
$\theta_{B_0}$
to zero.}. To see analytically what is occurring, we 
consider the intermediate and low $\tan \beta$ region, where the RGE can be 
analytically solved. One finds  for $B$ the result~\cite{aad}
\begin{equation}
B = B_0 -\frac{1}{2}(1 - D_0)  - \sum \Phi_i |m_{1/2}| e^{i \phi_i}
\end{equation}
where $D_0 = 1 - (m_t/ \sin \beta)^2 \stackrel{<}{\sim} 0.25$ and $\Phi_i =
O(1)$. As $\tan \beta$
gets large, the radiative 
breaking condition Eq.~(2) shows that $|B|$ gets small. Taking the imaginary 
part of Eq.~(27) gives
\begin{equation}
|B| \sin \theta_B = |B_0| \sin \theta_{B_0} 
- (1/2) (1 - D_0) |A_0| \sin \alpha_0 - \sum \Phi_i |m_{1/2}| \sin \phi_i    
\end{equation}
To the lowest approximation, since $|B|$ is small, one may then neglect the lhs
of Eq.~(28). Eq.~(28)
may 
then be viewed as an equation determining $\theta_{B_0}$ in terms of the other
GUT scale 
phases, and hence $\theta_{B_0}$ will in general be large if the other phases
are 
not all small (a result that is confirmed in \cite{aad} by detailed
calculation). 
However, the range of $\theta_{B_0}$ so that the experimental bound on $d_e$ is 
satisfied is then significantly reduced. Thus for  fixed values of 
$\alpha_0$  and $\phi_i$, Eq.~(28) gives as $\tan \beta$ gets large (i.e. $|B|$
becomes small)
\begin{equation}
\Delta \theta_{B_0} \cong (|B|/|B_0|) 
\Delta \theta_B \ll \Delta \theta_B                             
\end{equation}
Since we have already seen that $\Delta \theta_B$ is small (though not 
violating the fine tuning condition), one may expect that $\Delta \theta_{B_0}$
may need to be fine tuned, particularly for low $m_{1/2}$ where the SUSY mass 
spectrum is light. This is seen explicitly in 
Fig.~10 for $\tan \beta= 40$, $A_0 = 0$ 
with $\phi_1 = 0.9$, 1.2, 1.6, 2.3, and 2.6. We see that $R(\theta_{B_0})$
decreases 
as $\phi_1$ increases from 0.9 to 1.6 ($\cong \pi/2$) and then increases for
2.3 and 
2.6 where $\pi - \phi_1$ is decreasing. (The upper bound on $m_{1/2}$ arising
from 
the lower bound on $a_{\mu}$ decreases as $\phi_1$  moves into the second
quadrant in 
accord with Eqs.~(20,21).) We see that if one imposes the fine tuning 
constraint that $R(\theta_{B_0}) > 0.01$, large sections of the low $m_{1/2}$
region 
would be excluded, e. g. one would require $m_{1/2} > 540$ GeV for $\phi_1 =
1.6$. 
The fine tuning becomes more acute at lower values of $\tan \beta$. 
Thus Fig.~11 
shows $R(\theta_{B_0})$ as a function of $m_{1/2}$ for $\tan \beta = 15$, $A_0 =
0$ for $\phi_1 = 
0.9$ (upper curve) and $\phi_1 = 1.2$ (lower curve). Thus the fine tuning 
constraint would eliminate completely $\phi_1 = 1.2$ (and the entire region of
$\sim 0.4$ radians around $\pi/2$) and also restricts the other values of
$\phi_1$. 
Finally, we note that increasing $A_0$ generally increases the amount of fine 
tuning needed. Thus Fig.~12 shows $R(\theta_{B_0})$ for $\tan \beta = 40$,
$\phi_1 = 0.9$ 
for $A_0 = 0$ (upper curve) and $|A_0| = 2 m_{1/2}$, $\alpha_0 = 0.5$ (lower
curve). The 
entire $|A_0| = 2 m_{1/2}$ curve has $R(\theta_{B_0}) < 0.001$.

\begin{figure}[htb]
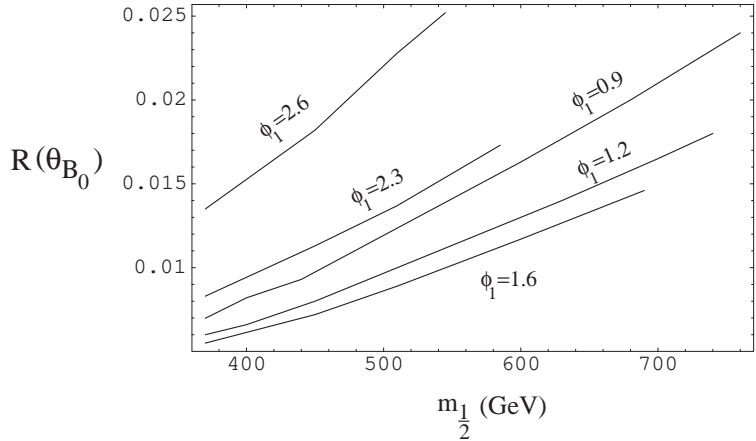

\centerline{ \DESepsf(edmt6.epsf width 10 cm) }
\caption {\label{fig10} 
$R(\theta_{B_0})$ as a function of $m_{1/2}$ for $\tan \beta = 40$, $A_0
= 0$ for (from 
bottom to top) $\phi_1 = 1.6$, 1.2, 0.9, 2.3 and 2.6.}
\end{figure}

\begin{figure}[htb]
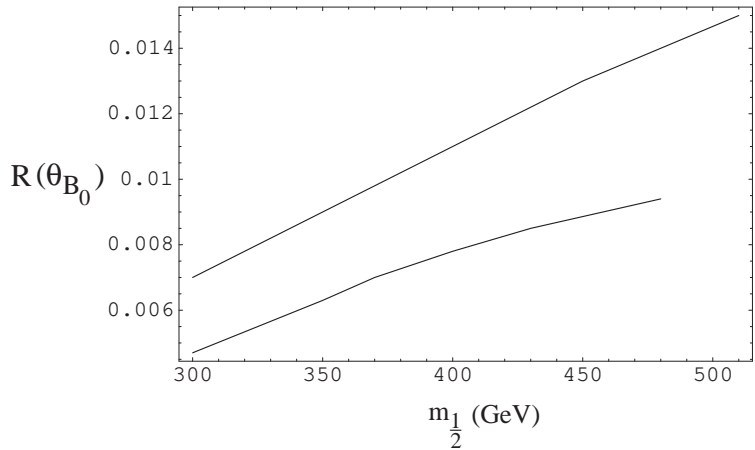

\centerline{ \DESepsf(edmt10.epsf width 10 cm) }
\caption {\label{fig11} 
$R(\theta_{B_0})$ as a function of $m_{1/2}$ for $\tan \beta = 15$, $A_0
= 0$ for $\phi_1 
= 0.9$ (upper curve) and $\phi_1 = 1.2$ (lower curve)}
\end{figure}

\begin{figure}[htb]
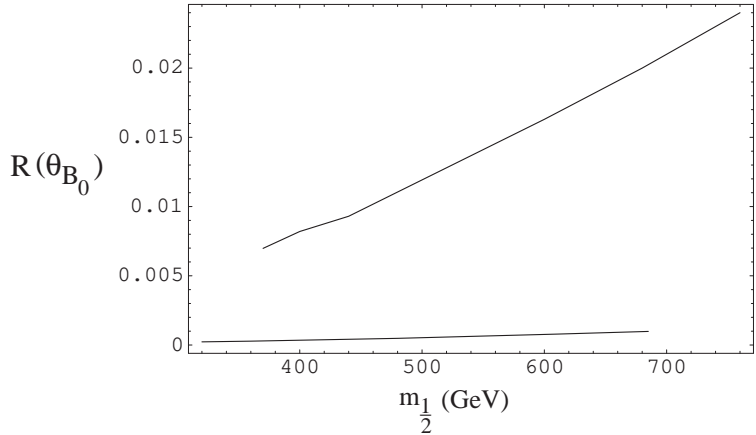

\centerline{ \DESepsf(edmt9.epsf width 10 cm) }
\caption {\label{fig12} 
$R(\theta_{B_0})$ as a function of $m_{1/2}$ for $\tan \beta = 40$,
$\phi_1 = 0.9$ for 
$A_0 = 0$ (upper curve) and $|A_0| = 2 m_{1/2}$, $\alpha_0 =0.5$ (lower curve).}
\end{figure}

\section{Conclusions}

We have examined here what SUSY CP violating phases are possible within the 
framework of SUGRA models, when the electron electric dipole moment 
experimental bound is imposed on the SUSY parameter space. In order to 
analyze this, we imposed simultaneously the accelerator bounds of the Higgs 
mass ($m_h > 114$ GeV) and the $b \rightarrow  s \gamma$ branching ratio, the
relic 
density bound for neutralino cold dark matter and the recent 2.6 std 
deviation of the muon magnetic moment from the Standard Model prediction. 
The different experimental constraints interact with each other. 
Thus the Higgs mass and $b \rightarrow s \gamma$ constraints put lower bounds on
the 
gaugino mass ($m_{1/2} \stackrel{>}{\sim} (300-400)$ GeV) which puts the relic
density analysis 
mostly in the region where stau-neutralino co-annihilation occurs. This 
closely fixes the scalar mass $m_0$ in terms of $m_{1/2}$ (for fixed $A_0$ and 
$\tan \beta$). The $a_{\mu}$ lower bound then puts an upper bound on $m_{1/2}$.
Thus the 
parameter space becomes highly constrained. One can estimate 
analytically, as was done in Sec. 2, the effects of turning on the CP 
violating phases. In fact if $d_e$ were zero, one could still have large CP 
violating phases present, with the Bino phase $\phi_1$ between 0 and $2\pi$. The
condition that $a_{\mu}$ be positive puts the $B$ phase $\theta_B$ in the first
and fourth quadrants with $|\theta_B| \sim 0.2 - 0.4$. 
The dependence of $\theta_B$ on $\tan \beta$ and 
$A_0$ was discussed in detail in Sec. 3.  It was seen that the phase $\phi_1$
acts 
to reduce the theoretical value of $a_{\mu}$ by a factor $Q < 1$, defined
in 
Eq.~(20, 21), and from the experimental lower bound on $a_{\mu}$ this reduces
the 
upper bound on $m_{1/2}$, limiting further the parameter space.

The relevant question, however, is whether the smallness of the 
experimental bound on $d_e$ requires an unreasonable amount of fine tuning of 
the phases. Using the parameter $R(\phi) = \Delta \phi/\phi_{\rm av}$, we find
at the 
electroweak scale, both $R(\theta_B)$ and $R(\phi_1)$ are small, i. e. a few 
percent. However, in general $R > 0.01$, and so no fine tuning below the $1 \%$ 
level is needed. However, at the GUT scale, this is not the case for 
$R(\theta_{B_0})$ for a significant part of the parameter space. Thus if we
were 
to exclude regions where $R(\theta_{B_0}) < 0.01$, then for $\tan \beta = 15$,
$A_0 = 0$, 
$\phi_1$ phases near $90^{\circ}$ ( i. e. $1.2 < \phi_1 < 2.0$) are completely
excluded 
(Fig.~11). The effect is reduced for higher $\tan \beta$. However, for example, 
for $\tan \beta = 40$, $A_0 = 0$ the condition $R > 0.01$ would eliminate
$m_{1/2} < 540$ GeV for $\phi_1 = 1.6$, and raise the lower bound on $m_{1/2}$ 
by a lesser amount for other values of $\phi_1$ (Fig.~10). Increasing $A_0$
decreases the 
value of $R$, making the fine tuning more serious, as can be seen in Fig.~12.

The fine tuning problem is thus on the verge of becoming quite acute. 
The experimental bound on $d_e$ is likely to decrease by 
a factor of three in the near future \cite{demille}. The effect this would have 
 is shown in Figs.~13 and 14. Fig.~13 shows 
$R(\theta_B)$ as a function of $m_{1/2}$ for $\tan \beta = 40$, $A_0 =0$,
$\phi_1 = 0.4$ (upper curve) (corresponding to $\theta_B\simeq0.1$)
 and $\phi_1 = 0.9$ (lower curve). The 
curves are what would occur if the experimental bound on $d_e$ were reduced by a
factor of 
three. Fig.~14 shows $R(\theta_{B_0})$ as a function of $m_{1/2}$ for $\tan
\beta = 15$, $A_0 
= 0$,  $\phi_1 = 0.3$ (upper curve) (corresponding to $\theta_B\simeq0.1$) and $\phi_1 = 0.9$ (lower curve), 
 if the bound is reduced by a factor of three. $\phi_1<0.4$ in Fig.~13 and 
 $\phi_1<0.3$ in Fig.~14 would have less fine tuning but correspond to  
 $\theta_B<0.1$.

\begin{figure}[htb]
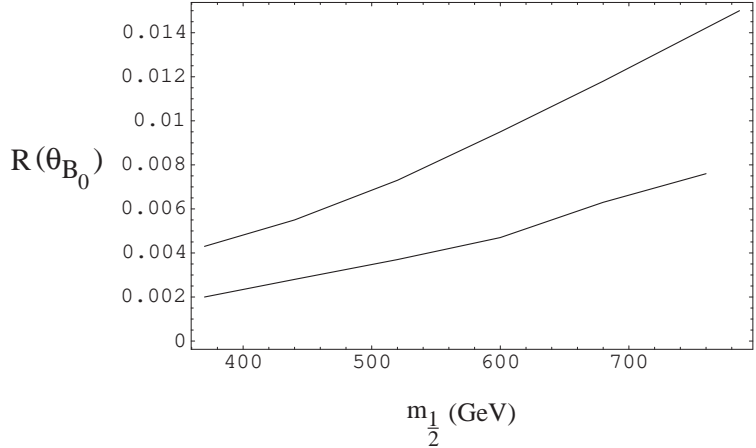

\centerline{ \DESepsf(edmt15.epsf width 10 cm) }
\caption {\label{fig13} 
$R(\theta_B)$ as a function of $m_{1/2}$ for $\tan \beta = 40$, $A_0 =
0$, $\phi_1 = 
0.9$ (lower curve) and $\phi_1 = 
0.4$ (upper curve)with the current experimental bound on $d_e$ reduced by a factor of three.}
\end{figure}

\begin{figure}[htb]
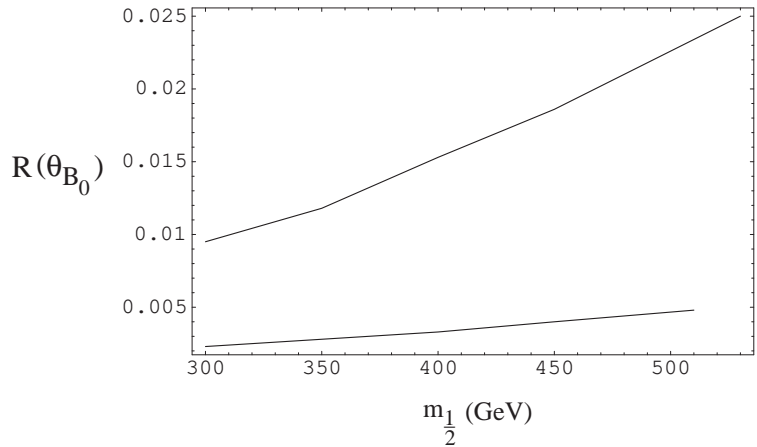

 \centerline{ \DESepsf(edmt14.epsf width 10 cm) }
\caption {\label{fig14}
$R(\theta_{B_0})$ as a function of $m_{1/2}$ for $\tan \beta = 15$, $A_0
=0$, $\phi_1 = 
0.9$ (lower curve) and $\phi_1 = 
0.3$ (upper curve)with the current experimental bound on $d_e$ 
reduced by a factor of three.}
\end{figure}

In a GUT theory, presumably the parameters at the GUT scale are the more 
fundamental ones, and fine tuning of these parameters would presumably 
represent a serious problem. Of course, what level of fine tuning one 
accepts is somewhat a matter of taste, and we view the criteria $R > 0.01$ as 
merely a bench mark for consideration. However, the above results show that 
any significant experimental reduction of $d_e$ would make the idea of large 
SUSY CP violating phases more difficult to maintain, as fine tuning would 
set in even at the electroweak scale unless the SUSY masses are 
heavy.

\bigskip
This work was supported in part by National Science Foundation grant No.
PHY-0070964. 

\appendix
\section{Derivations}
In this Appendix, we calculate the leading terms in $\tan \beta$ for the complex
amplitude $A$ of Eqs.~(5) and (6) needed to calculate $a_{\mu}$ and $d_e$. In
order to do this, it is necessary to diagonalize the various mass matrices
entering in the chargino and neutralino loops. These matrices depend on
the phases $\phi_1$, $\alpha_l$ and $\theta = \theta_{\mu} + \epsilon_1 +
\epsilon_2$ where $\alpha_l$ is the phase of $A_l$ and $\langle H_{1,2} \rangle
= v_{1,2} \; e^{i \epsilon_{1,2}}$ where 
$v_{1,2} = | \langle H_{1,2} \rangle |$.
Minimizing the effective potential with respect to $\epsilon_{1,2}$ determines
$\theta$ in terms of $\theta_B$, and one may then chose Higgs phases such that
$\epsilon_{1,2}=0$, as was done in Eq.~(1). With this choice of phases, the
chargino and neutralino mass matrices are
\begin{equation}
M_{\tilde{\chi}^{\pm}} = \left( \begin{array}{cc} \tilde{m}_2 & \sqrt{2} M_W
\sin \beta \\ \sqrt{2} M_W \cos \beta & \mu \end{array} \right)
\end{equation}

\begin{equation}
M_{\tilde{\chi}^{0}} = \left( \begin{array}{cccc} \tilde{m}_1 & 0 & a & b \\
0 & \tilde{m}_2 & c & d \\
a & c & 0 & -\mu \\
b & d & -\mu & 0 \end{array} \right)
\end{equation}
where $\tilde{m}_1 = |\tilde{m}_1| e^{i \phi_1}$, $\mu = |\mu| e^{i
\theta_{\mu}}$, $a = -M_Z \sin \theta_W \cos \beta$, $b=M_Z \sin \theta_W \sin
\beta$, $c = - a\cot \theta_W$, $d = -b \cot \theta_W$, and $\tilde{m}_2$ has
been chosen real and positive.

The slepton mass matrices have the form
\begin{equation}
M_{\tilde{l}}^2 = \left( \begin{array}{cc} m_{lLL}^2 & m_{lLR}^2 \\
m_{lLR}^{2 \; \ast} & m_{lRR}^2 \end{array} \right) 
\end{equation}
where $m_{lLR}^2 = m_l (A_l^{\ast} - \mu \tan \beta)$, $A_l = |A_l| e^{i
\alpha_l}$, and 
\begin{eqnarray}
m_{lLL}^2 &=& m_L^2 + m_l^2 - \frac{1}{2} (1 - 2 \sin^2 \theta_W) M_Z^2 \cos 2
\beta \\
m_{lRR}^2 &=& m_R^2 + m_l^2 - \sin^2 \theta_W M_Z^2 \cos 2 \beta 
\end{eqnarray}
The $m_{L,R}^2$ are obtained by running the RGE from the GUT scale to the
electroweak scale~\cite{rge}. In our analysis we will assume universal soft
breaking of
the $\tilde{\mu}$ and $\tilde{e}$ scalar masses ($m_0$) and universal cubic soft
breaking masses ($A_0$) at $M_G$. Since $m_e^2$ and $m_{\mu}^2$ are very small,
they can be neglected at the electroweak scale, i.e., $M_{\tilde{\mu}}^2 =
M_{\tilde{e}}^2$.

$M_{\tilde{\chi}^0}$ is a symmetric, complex matrix and can be diagonalized by a
unitary matrix $X$ according to $M_{\tilde{\chi}^0} X = X^{\ast}
M_{\tilde{\chi}^0}^{(D)}$ where 
\begin{equation}
M_{\tilde{\chi}^0}^{(D)} = {\rm diag}(m_{\tilde{\chi}_1^0}, m_{\tilde{\chi}_2^0},
m_{\tilde{\chi}_3^0}, m_{\tilde{\chi}_4^0})
\end{equation}
$M_{\tilde{l}}^2$ can be diagonalized by a hermitian matrix $D$ with
$M_{\tilde{l}}^2 D = D M_{\tilde{l}}^{2(D)}$ where $M_{\tilde{l}}^{2(D)} = {\rm
diag}(m_{\tilde{l}_1}^2, m_{\tilde{l}_2}^2)$. Finally one diagonalizes
$M_{\tilde{\chi}^{\pm}}$ by two unitary transformations $U$ and $V$ according to
$U^{\ast} M_{\tilde{\chi}^{\pm}} V^{\dag} = M_{\tilde{\chi}^{\pm}}^{(D)}$ where
$M_{\tilde{\chi}^{\pm}}^{(D)} = {\rm diag}(m_{\tilde{\chi}_1^{\pm}},
m_{\tilde{\chi}_2^{\pm}})$.

The amplitude $A$ can be divided into its chargino and neutralino parts: $A =
A(\tilde{\chi}^{\pm}) + A(\tilde{\chi}^{0})$. We follow the notation
of \cite{b34}
where one finds that 
\begin{equation}
A(\tilde{\chi}^{\pm}) = \frac{1}{\sqrt{2} M_W \cos \beta} \sum_i
\frac{1}{m_{\tilde{\chi}^{\pm}_i}} U_{i2}^{\ast} V_{i1}^{\ast} F_i
\end{equation}
and $F_i = F(m_{\tilde{\nu}}^2 / m_{\tilde{\chi}_i^{\pm}}^2 )$ with
\begin{equation}
F(x) = \frac{1 - 3x}{(1-x)^2} - \frac{2 x^2 \ln x}{(1-x)^3}
\end{equation}
Similarly
\begin{equation}
A(\tilde{\chi}^{0}) = \frac{1}{m_l} \sum_{k,j}
\frac{1}{m_{\tilde{\chi}^{0}_j}} \left( \eta_j^k G_{kj} + \frac{m_l}{6} X_j^k
H_{kj} \right) 
\end{equation}
where
\begin{eqnarray}
\eta_j^k &=& - \left[ \frac{1}{\sqrt{2}} ( \tan \theta_W X_{1j} + X_{2j} )
D_{1k}^{\ast} - \kappa_l X_{3j} D_{2k}^{\ast} \right] \nonumber \\
&& \times \left[ \sqrt{2} \tan \theta_W X_{1j} 
D_{2k} + \kappa_l X_{3j} D_{1k} \right], \\
X_j^k &=& \frac{1}{2} \tan^2 \theta_W |X_{1j}|^2 \left(|D_{1k}|^2 + 4 |D_{2k}|^2
\right) +  \frac{1}{2} |X_{2j}|^2 |D_{1k}|^2 \nonumber \\
&& + \tan \theta_W |D_{1k}|^2 X_{1j} X_{2j}^{\ast} + O(m_l)
\end{eqnarray}
and $\kappa_l = m_l / (\sqrt{2} M_W \cos \beta)$. The loop integrals are $G_{kj}
= G(m_{\tilde{l}_k}^2 / m_{\tilde{\chi}_j^0}^2 )$, $H_{kj}
= H(m_{\tilde{l}_k}^2 / m_{\tilde{\chi}_j^0}^2 )$ where
\begin{eqnarray}
G(x) = \frac{1+x}{(1-x)^2} + \frac{2x}{(1-x)^3} \ln x \\
H(x) = \frac{2 + 5x - x^2}{(1-x)^3} + \frac{6x}{(1-x)^4} \ln x
\end{eqnarray}
We need only keep the terms in $A$ independent of $m_l$ and thus can neglect the
$O(m_l)$ terms in Eq.~(A11). (Actually, $\eta_l^k$ begins linearly in $m_l$.)

We are interested in calculating only the leading terms in $\tan \beta$. We will
also do this in the limit $M_W^2 / |\mu|^2 \ll 1$ and $M_W^2 / |\tilde{m}_i|^2
\ll 1$ (which is valid for most of the mSUGRA parameter space). In that case one
has~\cite{b35}
\begin{equation}
m_{\tilde{\chi}_1^0} \cong |\tilde{m}_1|; \;\; m_{\tilde{\chi}_2^0} \cong
m_{\tilde{\chi}_1^{\pm}} \cong \tilde{m}_2; \;\; m_{\tilde{\chi}_{3,4}^0} \cong
m_{\tilde{\chi}_2^{\pm}} \cong |\mu|
\end{equation} 
We consider first the calculation of $A(\tilde{\chi}^{\pm})$. Since $V$
diagonalizes $M_{\tilde{\chi}^{\pm}}^{\dag} M_{\tilde{\chi}^{\pm}}$ and
$U^{\ast}$ diagonalizes $M_{\tilde{\chi}^{\pm}} M_{\tilde{\chi}^{\pm}}^{\dag}$
one finds for the leading terms
\begin{eqnarray}
U_{12}^{\ast} &\cong& - \frac{1}{\mu} \sqrt{2} M_W \sin \beta
\frac{|\mu|^2}{|\mu|^2-\tilde{m}_2^2} U_{11}^{\ast} \\
V_{21} &\cong& \frac{1}{\tilde{m}_2} \sqrt{2} M_W \sin \beta
\frac{\tilde{m}_2^2}{|\mu|^2-\tilde{m}_2^2} V_{22}
\end{eqnarray}
With an appropriate choices of phases one has to lowest order $U_{11} \cong 1
\cong U_{22}$, $V_{11} \cong 1$ and 
\begin{equation}  
V_{22} \cong e^{i \theta_{\mu}}
\end{equation} 
Hence inserting into Eq.~(A7) gives 
\begin{equation}
A(\tilde{\chi}^{\pm}) = - \frac{\tan \beta}{\tilde{m}_2 |\mu|} e^{-i
\theta_{\mu}} \left[ \frac{|\mu|^2}{|\mu|^2-\tilde{m}_2^2} F_1 -
\frac{\tilde{m}_2^2}{|\mu|^2-\tilde{m}_2^2} F_2 \right]
\end{equation}

To calculate the leading terms of $A(\tilde{\chi}^{0})$ it is useful to first
note the size of the matrix elements $X_{ij}$. Thus to zeroth order in $M_Z$
\begin{eqnarray}
X_{11} &\cong& e^{-\frac{i}{2} \phi_1}, \;\; X_{22} \cong 1 \\
X_{33} &\cong& - X_{43} \cong \frac{1}{\sqrt{2}} e^{-\frac{i}{2} \theta_{\mu}},
\;\; X_{34} \cong X_{44} \cong e^{\frac{\pi i}{2}} X_{33}
\end{eqnarray}
Also one has $X_{12}$, $X_{21}$ are $O(M_Z^2)$ (and hence negligible) while the
remaining components are $O(M_Z)$. To lowest order, the slepton mass eigenvalues
are $m_{\tilde{l}_1}^2 \cong m_{lRR}^2$, $m_{\tilde{l}_2}^2 \cong m_{lLL}^2$,
and the $D$ matrix has the form $D_{12} \cong 1 \cong D_{21}$ and 
\begin{equation}
D_{11} \cong - \frac{m_l (A_l^{\ast} - \mu \tan \beta)}{m_{\tilde{l}_2}^2 -
m_{\tilde{l}_1}^2} \cong - D_{22}^{\ast} 
\end{equation}

To illustrate the calculation of $A(\tilde{\chi}^0)$ we consider the leading
term when $k=1=j$. From Eq.~(A10), two terms contribute to $\eta_1^1$ for large
$\tan \beta$:
\begin{eqnarray}
\eta_1^1 & = & - \left( \frac{1}{\sqrt{2}} \tan \theta_W X_{11} D_{11}^{\ast}
\right) \left(\sqrt{2} \tan \theta_W X_{11} D_{21} \right) \nonumber \\
&& + \left( \kappa_l X_{31} D_{21}^{\ast} \right) 
\left(\sqrt{2} \tan \theta_W X_{11} D_{21} \right)
\end{eqnarray}
which evaluates to 
\begin{eqnarray}
\eta_1^1 & = & - \frac{m_l \tan^2 \theta_W \tan \beta}{|\mu|} \left[
\frac{|\mu|^2}{m_{\tilde{l}_2}^2 - m_{\tilde{l}_1}^2} - \frac{|\mu|^2}{|\mu|^2 -
|\tilde{m}_1|^2 } \right] e^{-i(\theta_{\mu} + \phi_1)} 
\end{eqnarray}
where we have used 
\begin{equation}
X_{31} \cong \frac{M_Z \sin \theta_W \sin \beta}{\mu} \frac{|\mu|^2}{|\mu|^2 -
|\tilde{m}_1|^2 } X_{11}
\end{equation}
(Note that $\eta_1^1$ is linear in $m_l$ and hence Eq.~(A9) is not singular as
$m_l \rightarrow 0 $.) In a similar fashion one can obtain all the leading terms
in Eq.~(A9). (The $X_j^k$ terms of Eq.~(A11) do not contribute.) The total
answer is 
\begin{eqnarray}
A(\tilde{\chi}^0) & \cong & 
- \frac{\tan^2 \theta_W \tan \beta}{|\tilde{m}_1| |\mu|}
\left[ \left\{ \left(\frac{|\mu|^2}{m_{\tilde{l}_2}^2 - m_{\tilde{l}_1}^2} - \frac{|\mu|^2}{|\mu|^2 -
|\tilde{m}_1|^2 } \right) G_{11} \right. \right. \nonumber \\
&& - \left. \left( \frac{|\mu|^2}{m_{\tilde{l}_2}^2 - m_{\tilde{l}_1}^2} - 
\frac{1}{2} \frac{|\mu|^2}{|\mu|^2 -
|\tilde{m}_1|^2 } \right) G_{21} \right\} e^{-i(\theta_{\mu} + \phi_1)}
\nonumber \\
&& - \frac{1}{2} \frac{1}{\tan^2 \theta_W} \frac{|\mu|^2}{|\mu|^2 -
\tilde{m}_2^2} \left( \frac{|\tilde{m}_1|}{\tilde{m}_2} G_{22} - \frac{1}{2}
\frac{|\tilde{m}_1| \tilde{m}_2}{|\mu|^2} G_{23} \right) e^{-i \theta_{\mu}}
\nonumber \\
&& + \left. \frac{1}{4} \frac{1}{\tan^2 \theta_W} \frac{|\tilde{m}_1|}{|\mu|}
\frac{|\mu|^2}{|\mu|^2 - \tilde{m}_2^2} G_{23} \right]
\end{eqnarray}
We note that a large amount of cancellation occurs in this regime: terms
proportional to $G_{24}$ have all canceled with part of the $G_{23}$ terms, and
the total $G_{14}$ contribution cancels with the $G_{13}$ terms. Note also that
the $A(\tilde{\chi}^0)$ depends separately on two phase combinations:
$\theta_{\mu} + \phi_1$ and $\theta_{\mu}$, though terms depending on
$\theta_{\mu} + \phi_1$ are generally larger.


\begin{thebibliography}{99}
\bibitem{CPexp}See e.g. BaBar Collaboration, hep-ex/0105073; 
BELLE Collaboration, \Journal{\NPA}{684}{704}{2001}; for a review see T. Hurth
et al., \Journal{\JPG}{27}{1277}{2001}. 
\bibitem{cancel}T. Falk and K. Olive, \Journal{\PLB}{375}{196}{1996}; 
T. Ibrahim and P. Nath \Journal{\PLB}{418}{98}{1998};
\Journal{\PRD}{57}{478}{1998}; Erratum-ibid.{\bf 58}, {019901} (1998); 
Erratum-ibid.{\bf 60}, {079903} (1999); Erratum-ibid.{\bf 60}, {119901} (1999). 
\bibitem{largephase}T. Falk and  K. Olive, \Journal{\PLB}{439}{71}{1998};
 M. Brhlik, G. Good and G. Kane, \Journal{\PRD}{59}{115004}{1999};
 M. Brhlik, L. Everett, G. Kane and J. Lykken, 
 \Journal{\PRL}{83}{2124}{1999}; \Journal{\PRD}{62}{035005}{2000};
 A. Bartl, T. Gajdosik, W. Porod, P. Stockinger and  H.
Stremnitzer, \Journal{\PRD}{60}{073003}{1999}; S. Pokorski, J. Rosiek and C. A. Savoy, 
\Journal{\NPB}{570}{81}{2000}. 
\bibitem{aad}E. Accomando, R. Arnowitt and B. Dutta,
\Journal{\PRD}{61}{075010}{2000}.
\bibitem{199HgEDM}T. Falk , K. Olive, M. Pospelov and R. Roiban, 
\Journal{\NPB}{560}{3}{1999}.
\bibitem{199HgEDM1}V. Barger, T. Falk, T. Han, J. Jiang, T. Li and T. Plehn,
 hep-ph/0101106;  
S. Abel, S. Khalil and  O. Lebedev, hep-ph/0103320.
\bibitem{BNL}H.N. Brown et.al., Muon (g-2) Collaboration,
\Journal{\PRL}{86}{2227}{2001}.
\bibitem{adhs}J. Feng and K. Matchev, \Journal{\PRL}{86}{3480}{2001}; 
U. Chattopadhyay and P. Nath, hep-ph/0102157; S. Komine, T. Moroi and
M. Yamaguchi, \Journal{\PLB}{506}{93}{2001}; \Journal{\PLB}{507}{224}{2001}; 
T. Ibrahim, U. Chattopadhyay and P. Nath, hep-ph/0102324; 
J. Hisano and K. Tobe, hep-ph/0102315;
J. Ellis, D.V. Nanopoulos and K. A. Olive, hep-ph/0102331; 
R. Arnowitt, B. Dutta, B. Hu and  Y. Santoso,
\Journal{\PLB}{505}{177}{2001};  S. Martin and J. Wells, hep-ph/0103067; 
H. Baer, C. Balazs, J. Ferrandis and X. Tata, hep-ph/0103280;
F. Richard, hep-ph/0104106; D. Carvalho, J. Ellis, 
M. Gomez and S. Lola, hep-ph/0103256; S. Baek, T. Goto, Y. Okada and 
K. Okumura, hep-ph/0104146; Y. Kim and M.  Nojiri, hep-ph/0104258. 
\bibitem{g2edm}T. Ibrahim, U. Chattopadhyay and P. Nath, hep-ph/0102324. 
\bibitem{sugra}A.H. Chamseddine, R. Arnowitt and P. Nath,
\Journal{\PRL}{49}{970}{1982}; R. Barbieri, S. Ferrara and C.A. Savoy,
\Journal{\PLB}{119}{343}{1982}; L. Hall, J. Lykken and S. Weinberg,
\Journal{\PRD}{27}{2359}{1983}; P. Nath, R. Arnowitt and A.H. Chamseddine,
\Journal{\NPB}{227}{121}{1983}.
\bibitem{anomaly}L. Randall and R. Sundrum, \Journal{\NPB}{557}{79}{1999}; G.
Giudice, M. Luty, H. Murayama and R. Rattazzi, JHEP {\bf 9812}, 027 (1998).
\bibitem{gauge}M. Dine and A. Nelson, \Journal{\PRD}{48}{1227}{1993}; M. Dine,
A. Nelson and Y. Shirman, \Journal{\PRD}{51}{1362}{1995}; M. Dine, A. Nelson, Y.
Nir and Y. Shirman, \Journal{\PRD}{53}{2658}{1996}; for a review see G. Giudice
and R. Rattazzi, \Journal{\PR}{322}{419}{1999}.
\bibitem{g-2ano}J. Feng and K. Matchev in ref.\cite{adhs}.
\bibitem{gmnoDM}T. Han and R. Hempfling, \Journal{\PLB}{415}{161}{1997}.
\bibitem{coan}R. Arnowitt, B. Dutta and Y. Santoso, hep-ph/0102181, {\em Nucl.
Phys.}
B, in press.
\bibitem{coan1} 
J Ellis, T. Falk, G. Ganis, K. Olive  and M. Srednicki; hep-ph/0102098; 
M. Gomez and J. Vergados,
hep-ph/0012020; M. Gomez, G. Lazarides and C. Pallis,
\Journal{\PRD}{61}{123512}{2000}; \Journal{\PLB}{487}{313}{2000}. 
\bibitem{b14}D.A. Kosower, L.M. Krauss and N. Sakai,
\Journal{\PLB}{133}{305}{1983}; T.C. Yuan, R. Arnowitt, A.H. Chamseddine and P.
Nath,
\Journal{\ZPC}{26}{407}{1984}.
\bibitem{b14a}J.L. Lopez, D.V. Nanopoulos and X. Wang,
\Journal{\PRD}{49}{366}{1994}; U. Chattopadhyay and P. Nath,
\Journal{\PRD}{53}{1648}{1996}; T. Moroi, \Journal{\PRD}{53}{6565}{1996};
Erratum-ibid, {\bf 56}, 4424 (1997).
M. Carena, G.F. Giudice and C.E.M. Wagner,
\Journal{\PLB}{390}{234}{1997}; 
T. Goto, Y. Okada and Y. Shimizu, hep-ph/9908499;
T. Blazek, hep-ph/9912460; G.C. Cho, K. Hagiwara and
M. Hayakawa, \Journal{\PLB}{478}{231}{2000}; T. Ibrahim and P. Nath,
\Journal{\PRD}{62}{015004}{2000}; M. Drees, Y. Kim,
T. Kobayashi and M. Nojiri, \Journal{\PRD}{63}{115009}{2001}.
\bibitem{b16}D. Demir, \Journal{\PRD}{60}{055006}{1999}.
\bibitem{nEDM}M. Pospelov and A. Ritz, \Journal{\PRD}{63}{073015}{2001}; M.
Hecht, C. Roberts and S. Schmidt, nucl-th/0101058. 
\bibitem{de}E.D. Commins et al., \Journal{\PRA}{50}{2960}{1994}.
\bibitem{mhexp}P. Igo-Kimenes, Talk presented at ICHEP 2000, Osaka, Japan, July
27 - August 2, 2000.
\bibitem{bsgexp}M. Alam et al., \Journal{\PRL}{74}{2885}{1995}.
\bibitem{bsgNLO}G. Degrassi, P. Gambino and G. Giudice,
\Journal{JHEP}{0012}{009}{2000}; M. Carena, D. Garcia, U. Nierste and C. Wagner,
\Journal{\PLB}{499}{141}{2001}.
\bibitem{aads}R. Rattazi and U. Sarid, \Journal{\PRD}{D53}{1553}{1996}; M.
Carena, M. Olechowski, S. Pokorski and C. Wagner, \Journal{\NPB}{426}{269}{1994}.
\bibitem{boomerang}C. Netterfield et al., astro-ph/0104460.
\bibitem{maxima}A.T. Lee et al., astro-ph/0104459 .
\bibitem{b30}R. Garisto and J. Wells, \Journal{\PRD}{55}{1611}{1997}. 
\bibitem{demille}Private communication with David DeMille.
\bibitem{rge}L. Ibanez and C. Lopez, \Journal{\NPB}{233}{511}{1984}.
\bibitem{b34}T. Ibrahim and P. Nath, \Journal{\PRD}{62}{015004}{2000}. 
\bibitem{b35}R. Arnowitt and P. Nath, \Journal{\PRL}{69}{725}{1992}.
\end{thebibliography}
\end{document}